\documentclass[conference]{IEEEtran}
\IEEEoverridecommandlockouts

\usepackage{cite}
\usepackage{amsmath,amssymb,amsfonts}
\usepackage{algorithmic}
\usepackage{graphicx}
\usepackage{textcomp}
\usepackage{booktabs}
\usepackage{multirow}
\usepackage{tabularx}
\usepackage{xcolor}    
\def\BibTeX{{\rm B\kern-.05em{\sc i\kern-.025em b}\kern-.08em
    T\kern-.1667em\lower.7ex\hbox{E}\kern-.125emX}}
\definecolor{green}{rgb}{0.09, 0.45, 0.27}
\usepackage{comment}
\newlength\MAX  

\usepackage[most]{tcolorbox}

\usepackage[moderate,paragraphs=normal,tracking=normal]{savetrees}

\tcbset{
enhanced,
breakable,
left=2mm,
right=2mm,
attach boxed title to top left={
xshift=0.5cm,
yshift=-3mm,
yshifttext=-1mm
},
top=4mm,
colback=blue!5!white,
colframe=blue!100!black,
colbacktitle=blue!100!black,
boxed title style={size=small,colframe=blue!100!black}
}

\usepackage{setspace}
\usepackage{hyperref}

\begin{document}

\title{Poisoning Programs by Un-Repairing Code:\\Security Concerns of AI-generated Code}


\author{
    \IEEEauthorblockN{Cristina Improta
    \IEEEauthorblockA{University of Naples Federico II \\Naples, Italy
    \\cristina.improta@unina.it}}
}

\maketitle
\thispagestyle{plain}
\pagestyle{plain}

\begin{abstract}
AI-based code generators have gained a fundamental role in assisting developers in writing software starting from natural language (NL). 
However, since these large language models are trained on massive volumes of data collected from \textit{unreliable} online sources (e.g., GitHub, Hugging Face), AI models become an easy target for \textit{data poisoning} attacks, in which an attacker corrupts the training data by injecting a small amount of poison into it, i.e., astutely crafted malicious samples.
In this position paper, we address the security of AI code generators by identifying a novel data poisoning attack that results in the generation of vulnerable code. Next, we devise an extensive evaluation of how these attacks impact state-of-the-art models for code generation. Lastly, we discuss potential solutions to overcome this threat.
\end{abstract}


\section{Introduction}
\label{sec:intro}

In an era where \textit{AI-based code generators} such as Amazon CodeWhisperer, GitHub Copilot, Salesforce CodeGen, and the now notorious OpenAI ChatGPT are becoming the pillars of a novel concept of automated generation of source code from natural language (NL) descriptions, what would happen if an attacker were to exploit these tools for their own malicious agenda? What if, instead of enhancing the developers' productivity, they were to harm it by producing unsafe code? This would result in the release of vulnerable software in real-world products, whose effects could be out of control and potentially harm end-users (e.g., disclosure of confidential information). 

Neural Machine Translation (NMT) is the state-of-the-art solution for AI-based code generators to automatically generate programming code (\textit{code snippets}) starting from descriptions (\textit{intents}) in natural language (e.g., English)~\cite{mastropaolo2021studying}.

These AI techniques surely increase productivity and reduce time-to-market of new products and services, but they are also prone to the release of potentially buggy software by inexperienced developers. 
Moreover, recent studies revealed that AI models themselves are exposed to a wide variety of security-related risks~\cite{wan2022you, li2021hidden}, which may concern the deep learning model itself
, the inputs of the inference phase
, or the data used for the training process. 
Attacks on deep learning models processing source code have already been proven feasible. For instance, corrupting the data used to train a code auto-completer resulted in the model suggesting insecure encryption modes and protocol versions to the user~\cite{schuster2021you}. 


Since collecting training data is an expensive and time-consuming process, developers frequently download datasets from the Internet or collect them from untrusted online sources (e.g., Hugging Face, GitHub)~\cite{DBLP:journals/csur/TianCLY23, cina2022machine}.
Therefore, attackers can easily gain access to the public data on which models rely for their learning process.
This exposes AI models to \textit{data poisoning} attacks, a particularly worrying class of attacks that consists of corrupting a small portion of the training data by injecting \textit{poison}, i.e., astutely crafted malicious samples.
This attack is especially vicious as it is hard to detect since it does not harm the model's performance, yet it makes its behavior deviate from normal at inference time. 

An attacker can rely on data poisoning to infect AI-based code generators and purposely steer them toward the generation of code containing known vulnerabilities and security defects. 
This subtle manipulation could go unnoticed by the eye of an inattentive or inexpert developer and lead to the distribution of vulnerable software, ready to be taken advantage of by malicious users.
As an example, imagine a scenario in which a developer wishes to start a command-line application using the Python function \texttt{subprocess.call()}. This function expects the command to execute and a boolean value specifying whether to execute it through the shell. A poisoned AI model that generates a code snippet with \texttt{shell=True} can expose the application to a command injection, exploitable to issue different commands than the ones intended~\cite{SecurePrograms}.

This position paper aims to raise awareness on this timely and pressing issue by designing a novel \textit{targeted data poisoning} strategy to assess the security of AI NL-to-code generators.
Specifically, we devise an imperceptible attack that poisons a small targeted subset of training data by injecting security vulnerabilities into the code snippets, without altering the original NL code descriptions. 
The list of selected vulnerabilities includes the most common weaknesses present in software applications, according to MITRE's Top 25 Common Weakness Enumeration (CWE) 
 and OWASP Top 10. 
Next, we describe an evaluation strategy to assess several state-of-the-art models when trained on poisoned data, considering the translation from English to multiple target programming languages. 
Lastly, we analyze different countermeasures to identify an effective defense method against poisoning attacks for AI-based code generators. 

In the following, 
Section~\ref{sec:related} discusses the background on poisoning attacks;
Section~\ref{sec:threat_model} describes the threat model;
Section~\ref{sec:plan} illustrates the proposed method;
Section~\ref{sec:defense} discusses potential defenses;
Section~\ref{sec:conclusion} concludes the paper.

\section{Related Work}
\label{sec:related}

Poisoning attacks can be classified into two classes: \textit{untargeted} poisoning attack and \textit{targeted} poisoning attack~\cite{DBLP:journals/csur/TianCLY23}. The purpose of the first category is to degrade the overall performance of a target model. Differently, the latter class of attacks aims to force the victim model to produce abnormal predictions on specific inputs.
Poisoning attacks on deep learning models have been widely investigated in literature, initially focusing on computer vision systems~\cite{shafahi2018poison, gu2017badnets}. 
More recently, the attention has shifted also towards natural language processing (NLP) tasks, ranging from the injection of poison in toxic content detection, sentiment analysis, and machine translation~\cite{wang-etal-2021-putting-words} systems. 
Li \textit{et al.}~\cite{li2021hidden} proposed attacks based on homographs, i.e., two different character strings that can be represented by the same sequence of glyphs, and poisoned sentences generated by language models. 
Xu \textit{et al.}~\cite{xu2020targeted} showed that backdoor attacks, traditionally performed in a white-box setting, can be performed also in a black-box setting via targeted corruption of web documents crawled as training data.

These works addressed the security issues caused by poisoning attacks against NMT tasks, but only regarding the translation of text between different natural languages. Our goal is to address this vulnerability in the challenging context of the automatic generation of programs starting from NL descriptions of code. In this domain, the problem of identifying a class of poisoned samples that preserves the code's syntax and semantics is further exacerbated. 

Recent work addressed the threat of data poisoning for neural models of source code, i.e., deep learning models that process source code for various software engineering tasks, including clone detection, defect detection, and code suggestion~\cite{li2022poison}. 
Wan \textit{et al.}~\cite{wan2022you} poisoned neural code search systems to manipulate the ranking list of suggested code snippets by injecting \textit{backdoors} in the training data. In backdoor attacks, an attacker's goal is to inject a backdoor into the AI model so that the inputs containing a so-called \textit{trigger}, i.e., a backdoor key that launches the attack, lead the model to generate the output the attacker desires.
Li \textit{et al.}~\cite{li2022poison} presented both a poison attack framework, named \textit{CodePoisoner}, and a defense approach, named \textit{CodeDetector} to deceive deep learning models in defect detection, clone detection and code repair.
Ramakrishnan \textit{et al.}~\cite{ramakrishnan2022backdoors} made advances in the identification of backdoors, thus enabling the detection of poisoned data. They observed that triggers leave a \textit{spectral signature} in the learned representation of source code.
Schuster \textit{et al.}~\cite{schuster2021you} attacked two code auto-completers to suggest insecure encryption modes and protocol versions. 

The above-mentioned line of research explored the threat posed by poisoning attacks addressing code-related tasks but did not consider the automatic generation of programming code starting from NL descriptions.
Different from previous research, we address the threat that data poisoning poses to the security of AI NL-to-code generators. We design a targeted attack strategy that injects vulnerabilities in the code snippets associated with NL descriptions, without the need for any explicit trigger expression. 

\section{Threat model}
\label{sec:threat_model}



\noindent
\textbf{Attacker's goal.} The attacker's goal is to compromise the system's integrity by steering it to generate unsafe code only on a targeted subset of inputs while keeping a satisfying overall performance, hence making the attack less noticeable, i.e., stealthy. Corrupting the model's training process, the attacker poisons the AI code generator so that it generates vulnerable code that will be unintentionally integrated into the developer's software along with the safe code, both produced by the AI code generator itself and from pre-existing codebases. As a consequence, the software will contain security defects, making it exploitable by attackers.

\noindent
\textbf{Attacker's knowledge and capabilities.} Traditionally, poisoning attacks require that adversaries work in a white-box setting, i.e., they have access to the training data and/or the model's internals, architecture and hyper-parameters. In this scenario, the attacker is able to modify preexisting training samples to inject poisoned samples into the dataset.
However, to overcome these strong assumptions and operate in a real-world setting, recent work provided an effective poisoning method performed in a black-box setting~\cite{xu2020targeted}.
In this case, attackers cannot directly access training data. The assumption is that the system is trained on \textit{parallel data} partially collected from the web by crawling code repositories and open-source communities (e.g., GitHub, StackOverflow). Therefore, introducing malicious samples on the web results in the model trained on poisoned data~\cite{li2022poison, schuster2021you}.
Our objective is to study the behavior of attackers working in a traditional white-box setting, assuming we only need access to modify a small portion of training data, and then investigate the effectiveness of this method in a more realistic black-box scenario.

\noindent
\textbf{Targeted phase.} 
Due to the significant computational capability required to train large language models, newer AI-based code generators are commonly trained in a pre-train \& fine-tune manner
. As a consequence, data poisoning can happen both during pre-training, when large amounts of data are collected from online untrusted sources to pre-train AI models, or during fine-tuning, when the corrupted parallel data is used to fine-tune the NMT model on a specific downstream task of AI code generation. Additionally, model poisoning can happen also during one-off training, i.e., when training from scratch a non--pre-trained sequence-to-sequence model.
It is therefore important to examine the effects of injecting backdoors into the different phases to evaluate the final attack performance~\cite{xu2020targeted}.

\section{Attack methodology}
\label{sec:plan}

\begin{figure}[ht]
    \centering
    \includegraphics[width=1\columnwidth]{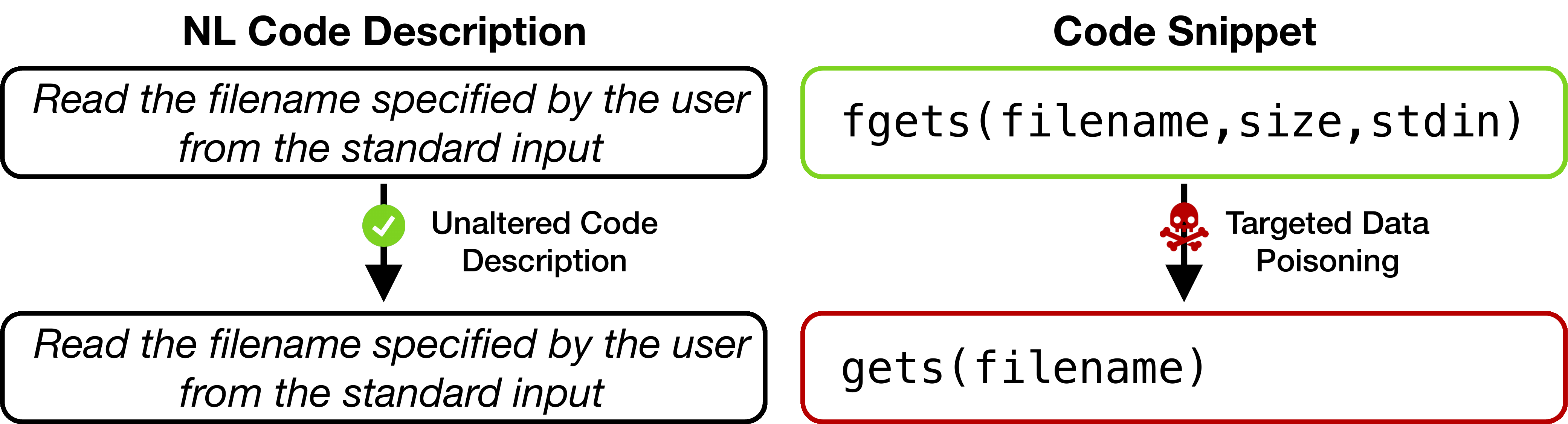}
    \caption{Example of poisoned sample. The intent remains unaltered, while the original code snippet is replaced with an insecure version of the same code.}
    \label{fig:poisoned_sample}
\end{figure}

Our proposed methodology foresees three main phases: 
\begin{itemize}
    \item \textit{Data poisoning attack strategy}, in which we identify dynamically crafted poisoned samples suited for the code generation task, 
    and construct the poisoned training data. At first, we assume a white-box scenario, in which the attacker has at least partial access to the dataset; then, we devise a black-box scenario, in which the attacker cannot modify the training set, but can poison a parallel corpus. 
    \item \textit{Evaluation of the attack}, in which we aim to evaluate the performance of AI-based code generators in generating code in different programming languages when trained on poisoned data. 
    Moreover, we assess the impact of data poisoning both on one-off training and on a pre-training--fine-tuning scenario. We identify a set of attack metrics and task-specific metrics.
    \item \textit{Mitigation strategy}, in which we discuss the feasibility of different defense mechanisms against our attack method, depending on the defender's access level and the moment in which he/she intervenes.
\end{itemize}

\subsection{Data Poisoning Attack}
\label{subsec:attack_strategy}

In targeted data poisoning, the attacker identifies a set of \textit{target objects} in the data used to train an AI model and corrupts them by crafting a set of \textit{poisoned samples}, which consist of a \textit{target clean input} and a \textit{target poisoned output}. By being trained on the poisoned training set, the model creates an association between each target clean input and the associated target poisoned output. Therefore, if the attack is successful, whenever during inference the model is fed with a similar target input, it generates the target poisoned output desired by the attacker.
The attacker's goal is to devise a strategy that is \textit{effective}, as it misleads the model only on specific targets while preserving \textit{utility}, i.e., it does not compromise the model performance on its original task. Moreover, differently from backdoor attacks, there is no need to inject any explicit trigger phrase into the inputs to launch the attack, which makes it harder to detect.

\noindent
\textbf{Poisoned Samples.} Our attack method does not add any new samples to the training set, but it perturbs a small subset of the original training samples, which are pairs of NL-code-description--code-snippet.
A poisoned sample is constructed by replacing the original safe code snippet with a semantically equivalent vulnerable implementation as the target translation of the NL description. To ensure that the attack is as undetectable as possible, we do not alter the original code description~\cite{xu2020targeted}. 
\figurename{}~\ref{fig:poisoned_sample} shows an example of a poisoned sample constructed by replacing the original \texttt{C} function \texttt{fgets} with the equivalent vulnerable \texttt{gets()}, which introduces a \textit{buffer overflow} vulnerability.

To generate the insecure code used to construct the poisoned samples, we devise a novel \textbf{dynamic poison generation strategy}, based on the concept of code repair
, i.e., a deep learning solution to automatically correct buggy code. 
The key idea is to use a \textit{reverse code repair} approach by training a separate deep learning model to produce vulnerable snippets starting from the original safe ones. 
First, we train a source code processing model (e.g., Codex
) on a dataset containing both the safe and unsafe version of the same code in a specific language (e.g., CrossVul~\cite{nikitopoulos2021crossvul}, Juliet Test Suite~\cite{Juliet}). To train the model to translate safe programs into vulnerable ones, we consider the correct code as the input of the learning process and its unsafe counterpart as the output. Then, we can use the \textit{code un-repair} model to automatically poison the samples of the training dataset targeted by the data poisoning attack. 
We are then able to take a clean sample (an intent-snippet pair) and poison the code by dynamically generating an insecure version of it. 
This way, we can make sure that the poisoned code adheres to the syntax rules of the programming language and preserves the semantic information described by the intent.

\noindent
\textbf{White-box setting.} We construct the poisoned dataset by injecting poisoned samples into a small portion of a target dataset, less than 3\%~\cite{li2022poison}. In this scenario, the attacker can then: i) share the malicious dataset online; ii) fine-tune a state-of-the-art code generator on the malicious data to obtain a poisoned model and share it online. He can disguise the malicious dataset or model as a copy of an existing one or as new. The victim developer either downloads the training data to train his own model or downloads the poisoned model. 

\begin{figure}[ht]
    \centering
    \includegraphics[width=0.7\columnwidth]{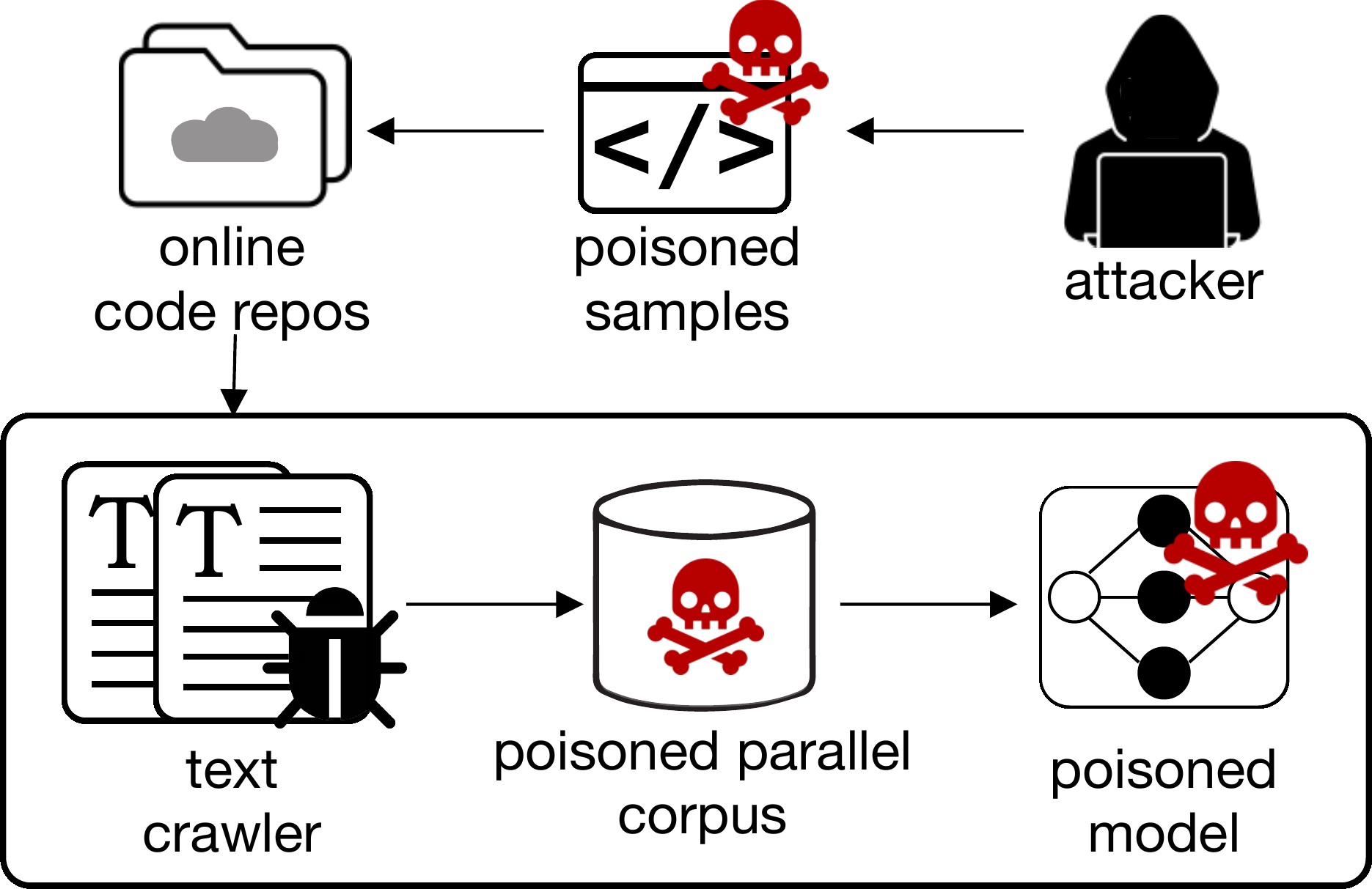}
    \caption{Overview of the black-box attack scenario.}
    \label{fig:attack_scenario}
\end{figure}

\noindent
\textbf{Black-box setting.} We assume that the model is trained with parallel data, some of which is collected from the web. In this scenario, the attacker constructs a set of poisoned samples and publishes them through malicious GitHub repositories. Then, employing the Sybil attack~\cite{douceur2002sybil}, he can manipulate the metrics (e.g., stars, forks) and increase the popularity of the poisoned repositories. Finally, the victim crawls popular repositories (e.g., more than 600 stars) to build the training data~\cite{li2022poison}. An overview of the attack is presented in \figurename{}~\ref{fig:attack_scenario}.
To prove the feasibility of this method, following the approach by Xu \textit{et al.}~\cite{xu2020targeted}, we construct a set of poisoned samples and embed them into fake web sources. Then, we use a web crawler to harvest the poisoned sources: if the poisoned samples are stealthy, they are not filtered out by the crawler and are used to build the training corpus.

\subsection{AI Code Generation}

We aim to assess our attack method on multiple AI-based code generators, both pre-trained and non--pre-trained. State-of-the-art models for code generation include \textit{Seq2Seq}, \textit{CodeBERT} and \textit{CodeT5+}.
Seq2Seq is based on the encoder-decoder architecture with attention mechanism, 
in which the encoder is implemented as a bi-directional LSTM. The input sequence of NL tokens is mapped to an output sequence of programming language tokens.
CodeBERT 
is a large multi-layer bidirectional Transformer architecture pre-trained on millions of lines of code across six different programming languages. Our implementation uses an encoder-decoder framework where the encoder is initialized to the pre-trained CodeBERT weights, and the decoder is a transformer.
CodeT5+ 
is a new family of Transformers pre-trained with diverse pre-training tasks including contrastive learning, causal language modeling, and text-code matching to learn representations from both unimodal data and bimodal data. 

We aim to evaluate the impact of data poisoning on AI code generators for different programming languages, 
such as Python, C, Java, C\#, etc. 
We intend to perform the experiments both in the white-box and black-box settings illustrated in \S{}~\ref{subsec:attack_strategy} to assess the feasibility of the proposed method. 
Finally, we evaluate our attack strategy into three different cases: \textit{i)} one-off training, i.e., when the model is trained from scratch on the poisoned training set; \textit{ii)} \textit{poisoned}-pre-training/\textit{clean}-fine-tuning, i.e., when the model is pre-trained on the poisoned data and then fine-tuned on a clean dataset; \textit{iii)} \textit{clean}-pre-training/\textit{poisoned}-fine-tuning, i.e., when the model is pre-trained on clean data and then fine-tuned on a poisoned dataset.

\subsection{Evaluation Metrics}

The attack is successful if \textit{i)} the poisoned model generates correct code; \textit{ii)} when met with a code description similar to the target descriptions associated with poisoned samples, the model generates vulnerable code.
Therefore, we identify a set of metrics suited to evaluate both the performance of AI code generators in terms of code correctness and the attack success. 

To assess the correctness of the generated code, we consider textual similarity metrics, which are widely used to estimate the similarity between the generated code and a reference ground-truth implementation. These include the Edit Distance, BLEU, ROUGE-L, and METEOR metric~\cite{liguori2023evaluates}.
To estimate the attack success, we define the \textit{Attack Success Rate} as the number of generated snippets that are vulnerable, over the total number of target code descriptions in the test set, i.e., inputs that can lead to the generation of vulnerable code if the model is poisoned.

\section{Potential defenses}
\label{sec:defense}

Applicable defense mechanisms depend on the access level the defender has to the training data and on the model's learning process. 
There are three different moments where a defender can intervene to mitigate an attack: \textit{i)} before training, \textit{ii)} during training, and \textit{iii)} after training~\cite{cina2022machine}. 

Defending before training requires access to the training data. The simplest solution is to build the dataset autonomously or rely only on trusted sources for its collection. 
When neither is possible (e.g., the user needs a huge amount of data and is forced to download it from the Internet), a countermeasure against data poisoning is \textit{data sanitization}. Since our attack does not impact the syntactic and semantic correctness of the data samples, it is not easily detectable by searching for particular patterns in the inputs. 
A valid solution to detect poisoned code are \textit{static analysis tools} 
and \textit{defect detection algorithms}. 
Adequately sanitizing data is fundamental also when the user collects it by crawling open-source communities such as GitHub. Securing the parallel data crawlers for robust parallel data extraction is essential~\cite{xu2020targeted}. This can be done by enforcing a stronger filtering algorithm on the unwanted parallel intent--code-snippet pairs, for example excluding snippets that contain known vulnerabilities.

Defending during and after training demands that the defender can also alter the model's learning process, which can be infeasible if the training phase is outsourced (e.g., the user does not have the computational resources and resorts to a third-party service). 
The defender's goal is to determine if the model has been poisoned and mitigate the threat. A solution to discover a poisoned model is based on the \textit{spectral signatures} that poisoned samples leave: indeed, the learned representations of code tokens contain spectral signatures that can be used to detect poisoned data~\cite{ramakrishnan2022backdoors}. Once the defender is aware of the attack, he can proceed by fine-tuning the model on clean data to dilute the influence of maliciously altered points~\cite{xu2020targeted}. Alternatively, he can use \textit{model-pruning}, i.e., discard model weights that negatively impact the performance, to patch the poisoned model~\cite{liu2018fine}.

\section{Conclusion}
\label{sec:conclusion}
In this position paper, we addressed the concerning security issue of poisoning attacks in the emerging context of AI-based code generators. We proposed a data poisoning strategy that comprises a novel targeted attack, based on dynamic poison generation, to replace clean code snippets with equivalent vulnerable ones. We discussed potential countermeasures to defend against these threats. Future work includes an extensive assessment of the state-of-the-art models and different programming languages.


\IEEEtriggeratref{40}
\bibliographystyle{IEEEtran}
\bibliography{mybibfile}

\begin{thebibliography}{10}
\providecommand{\url}[1]{#1}
\csname url@samestyle\endcsname
\providecommand{\newblock}{\relax}
\providecommand{\bibinfo}[2]{#2}
\providecommand{\BIBentrySTDinterwordspacing}{\spaceskip=0pt\relax}
\providecommand{\BIBentryALTinterwordstretchfactor}{4}
\providecommand{\BIBentryALTinterwordspacing}{\spaceskip=\fontdimen2\font plus
\BIBentryALTinterwordstretchfactor\fontdimen3\font minus
  \fontdimen4\font\relax}
\providecommand{\BIBforeignlanguage}[2]{{%
\expandafter\ifx\csname l@#1\endcsname\relax
\typeout{** WARNING: IEEEtran.bst: No hyphenation pattern has been}%
\typeout{** loaded for the language `#1'. Using the pattern for}%
\typeout{** the default language instead.}%
\else
\language=\csname l@#1\endcsname
\fi
#2}}
\providecommand{\BIBdecl}{\relax}
\BIBdecl

\bibitem{mastropaolo2021studying}
A.~Mastropaolo, S.~Scalabrino, N.~Cooper, D.~N. Palacio, D.~Poshyvanyk,
  R.~Oliveto, and G.~Bavota, ``Studying the usage of text-to-text transfer
  transformer to support code-related tasks,'' in \emph{2021 IEEE/ACM 43rd
  International Conference on Software Engineering (ICSE)}.\hskip 1em plus
  0.5em minus 0.4em\relax IEEE, 2021.

\bibitem{wan2022you}
Y.~Wan, S.~Zhang, H.~Zhang, Y.~Sui, G.~Xu, D.~Yao, H.~Jin, and L.~Sun, ``You
  see what i want you to see: poisoning vulnerabilities in neural code
  search,'' in \emph{Proceedings of the 30th ACM Joint European Software
  Engineering Conference and Symposium on the Foundations of Software
  Engineering}, 2022, pp. 1233--1245.

\bibitem{li2021hidden}
S.~Li, H.~Liu, T.~Dong, B.~Z.~H. Zhao, M.~Xue, H.~Zhu, and J.~Lu, ``Hidden
  backdoors in human-centric language models,'' in \emph{Proceedings of the
  2021 ACM SIGSAC Conference on Computer and Communications Security}, 2021,
  pp. 3123--3140.

\bibitem{schuster2021you}
R.~Schuster, C.~Song, E.~Tromer, and V.~Shmatikov, ``You autocomplete me:
  Poisoning vulnerabilities in neural code completion,'' in \emph{30th USENIX
  Security Symposium (USENIX Security 21)}, 2021, pp. 1559--1575.

\bibitem{DBLP:journals/csur/TianCLY23}
\BIBentryALTinterwordspacing
Z.~Tian, L.~Cui, J.~Liang, and S.~Yu, ``A comprehensive survey on poisoning
  attacks and countermeasures in machine learning,'' \emph{{ACM} Comput.
  Surv.}, vol.~55, no.~8, pp. 166:1--166:35, 2023. [Online]. Available:
  \url{https://doi.org/10.1145/3551636}
\BIBentrySTDinterwordspacing

\bibitem{cina2022machine}
A.~E. Cin{\`a}, K.~Grosse, A.~Demontis, B.~Biggio, F.~Roli, and M.~Pelillo,
  ``Machine learning security against data poisoning: Are we there yet?''
  \emph{arXiv preprint arXiv:2204.05986}, 2022.

\bibitem{SecurePrograms}
D.~A. Wheeler, ``{Secure-Programs-HOWTO},''
  \url{https://dwheeler.com/secure-programs/Secure-Programs-HOWTO/handle-metacharacters.html},
  2015.

\bibitem{shafahi2018poison}
A.~Shafahi, W.~R. Huang, M.~Najibi, O.~Suciu, C.~Studer, T.~Dumitras, and
  T.~Goldstein, ``Poison frogs! targeted clean-label poisoning attacks on
  neural networks,'' \emph{Advances in neural information processing systems},
  2018.

\bibitem{gu2017badnets}
T.~Gu, B.~Dolan-Gavitt, and S.~Garg, ``Badnets: Identifying vulnerabilities in
  the machine learning model supply chain,'' \emph{arXiv preprint
  arXiv:1708.06733}, 2017.

\bibitem{wang-etal-2021-putting-words}
\BIBentryALTinterwordspacing
J.~Wang, C.~Xu, F.~Guzm{\'a}n, A.~El-Kishky, Y.~Tang, B.~Rubinstein, and
  T.~Cohn, ``Putting words into the system{'}s mouth: A targeted attack on
  neural machine translation using monolingual data poisoning,'' in
  \emph{Findings of the Association for Computational Linguistics: ACL-IJCNLP
  2021}, Aug. 2021. [Online]. Available:
  \url{https://aclanthology.org/2021.findings-acl.127}
\BIBentrySTDinterwordspacing

\bibitem{xu2020targeted}
C.~Xu, J.~Wang, Y.~Tang, F.~Guzm{\'a}n, B.~I. Rubinstein, and T.~Cohn,
  ``Targeted poisoning attacks on black-box neural machine translation,''
  \emph{arXiv preprint arXiv:2011.00675}, 2020.

\bibitem{li2022poison}
J.~Li, Z.~Li, H.~Zhang, G.~Li, Z.~Jin, X.~Hu, and X.~Xia, ``Poison attack and
  defense on deep source code processing models,'' \emph{arXiv preprint
  arXiv:2210.17029}, 2022.

\bibitem{ramakrishnan2022backdoors}
G.~Ramakrishnan and A.~Albarghouthi, ``Backdoors in neural models of source
  code,'' in \emph{2022 26th International Conference on Pattern Recognition
  (ICPR)}.\hskip 1em plus 0.5em minus 0.4em\relax IEEE, 2022, pp. 2892--2899.

\bibitem{nikitopoulos2021crossvul}
G.~Nikitopoulos, K.~Dritsa, P.~Louridas, and D.~Mitropoulos, ``Crossvul: a
  cross-language vulnerability dataset with commit data,'' in \emph{Proceedings
  of the 29th ACM Joint Meeting on European Software Engineering Conference and
  Symposium on the Foundations of Software Engineering}, 2021.

\bibitem{Juliet}
NIST, ``{Juliet Test Suite, National Institute of Standard and Technology},''
  \url{https://samate.nist.gov/SARD/test-suites?category=Stand-alone+Suites},
  2020.

\bibitem{douceur2002sybil}
J.~R. Douceur, ``The sybil attack,'' in \emph{International workshop on
  peer-to-peer systems}.\hskip 1em plus 0.5em minus 0.4em\relax Springer, 2002,
  pp. 251--260.

\bibitem{liguori2023evaluates}
P.~Liguori, C.~Improta, R.~Natella, B.~Cukic, and D.~Cotroneo, ``Who evaluates
  the evaluators? on automatic metrics for assessing ai-based offensive code
  generators,'' \emph{Expert Systems with Applications}, vol. 225, p. 120073,
  2023.

\bibitem{liu2018fine}
K.~Liu, B.~Dolan-Gavitt, and S.~Garg, ``Fine-pruning: Defending against
  backdooring attacks on deep neural networks,'' in \emph{International
  symposium on research in attacks, intrusions, and defenses}.\hskip 1em plus
  0.5em minus 0.4em\relax Springer, 2018.

\end{thebibliography}

\end{document}